
\input phyzzx
\hoffset=0.2truein
\voffset=0.1truein
\hsize=6truein
\def\TITLEPAGE{\frontpagetrue}
\def\CALT#1{\hbox to\hsize{\tenpoint \baselineskip=12pt
        \hfil\vtop{
        \hbox{\strut CALT-68-#1}}}}

\def\CALTECH{
        \address{California Institute of Technology,
Pasadena, CA 91125}}

\def\AUTHOR#1{\vskip .2in \centerline{#1}}

\def\ABSTRACT#1{\vskip .2in \vfil \centerline{\twelvepoint
\bf Abstract}
        #1 \vfil}
\def\ENDTITLEPAGE{\vfil\eject\pageno=1}

\tolerance=10000
\hfuzz=5pt
\TITLEPAGE
\CALT{1886}
\bigskip           
\titlestyle {Path Integrals, Density Matrices, \break and Information
Flow with Closed Timelike Curves\foot{Work supported in part by the U.S. Dept.
of Energy
under Grant No. DE-FG03-92-ER40701.}}
\AUTHOR{H. David Politzer}
\CALTECH
\ABSTRACT{Two formulations of quantum mechanics, inequivalent in the
presence of closed timelike curves, are studied in the context of a
soluable system.  It illustrates how quantum field nonlinearities lead
to a breakdown of unitarity, causality, and superposition using a path
integral.  Deutsch's density matrix approach is causal but typically
destroys coherence.  For each of these formulations I demonstrate that
there are yet further alternatives in prescribing the handling of
information flow (inequivalent to previous analyses) that have
implications for any system in which unitarity or coherence are not
preserved.}

\ENDTITLEPAGE

\eject

Quantum mechanics has been suggested$^{[1,2]}$ as a possible means of
resolving some of the classic paradoxes of time travel.  The basic idea
is to require some sort of consistency around closed timelike curves
(CTC's).  As it is not yet known whether or not a compact, bounded region of
CTC's can arise$^{[3]}$ (due to gravitational interactions of quantum
mechanical matter), one possible line of inquiry is: given such a
spacetime, can it support any consistent, non-trivial mechanics?  Two
distinct generalizations of quantum mechanics have been proposed.  One
is defined by a coherent, action-weighted sum over all single-valued
histories defined on the spacetime.$^{[1]}$  The other is a linear time
evolution of a density matrix subject to certain periodic boundary
conditions around the CTC's.  The resulting mechanics of the two schemes
are quite
different, certainly peculiar, but in no apparent way self-inconsistent.

One purpose of this paper is to analyze  in some detail a system that is
simple enough to be solved exactly in both formulations but rich enough
to exhibit various striking phenomena.  In fact, intermediate steps in
the calculations involve nothing more complicated than the interaction of two
half-integer spins.  However, I will indicate the parallels between this
system and any arbitrary, interacting, quantum field theory.

A second goal is to illuminate issues regarding the flow of information
in circumstances where unitarity or coherence are not preserved by the
dynamics.\foot{Black holes are thought by some to be more plausible
examples of such systems than time machines.}  After establishing how
pure initial states evolve according to the two proposed generalizations
of quantum mechanics, I will show that the handling of initial mixed
states is not unambiguous in either formulation.  In particular, I will
argue for an implementation of the density matrix mechanics (motivated
by fairly classical notions of ensembles and probability) that it is
inequivalent to the formulation proposed in Reference 2.  In the path
integral mechanics, I show that the density matrix does not encode all
that can be known about a mixed state; however, the dynamics preserves
what is known about the system, although traditional entropy is not a
good measure of information.

In the context of the path integral approach, one new result presented
here is
that the non-unitarity of evolution from before to after a compact epoch
of CTC's -- hitherto identified in perturbation theory$^{[4-7]}$ --
persists in the exact solution of a non-linear theory.  (Previously, the
only interacting models solved exactly exhibited no
non-unitarity.$^{[7]}$)  Although this phenomenon has been identified
earlier, it may be of interest to see how, in a very simple context,
CTC's lead to non-unitary amplitudes using an action that would
otherwise have preserved unitarity in the absence of CTC's.
It may also be helpful to have an
explicit example in which to implement Hartle's general analysis of
non-unitary evolution,$^{[8]}$ which amounts to renormalizations to
preserve a probabilistic interpretation.  I present, in passing, an
additional argument to those given by Hartle as to the necessity of
following histories to the future of all CTC's.  A remarkable
consequence of non-unitarity and the consideration of full histories is
that experiments completed before the CTC epoch are sensitive to the
existence of CTC's in their future.  In particular, such experiments
violate the quantum superposition principle.

Deutsch's density matrix mechanics was originally presented in an even
more abstract context than the model analyzed here.  So certain
particulars, which he did not address, need to be established.  The
hallmark of the density matrix approach is that pure states can evolve
into mixed states after traversing a compact CTC epoch.  Physics before
that epoch is the same whether the CTC's come into existence or not,
thus evading the acausality that arises in the path integral
formulation.

Comparing the two approaches, we will see that there are interactions
which lead to non-unitary evolution from the path integral but preserve
the coherence of any initial pure state under density matrix evolution.
Conversely, there are interactions that do not preserve coherence using
density matrices but are totally unitary in the path integral sense.
More generally, these examples (which involve only four-by-four matrix
Hamiltonians) offer some insight into possible distinctions between
complex and simple quantum systems.

\noindent {\bf The Model System}

I will describe the model system (which is really only two coupled
spins) as an abbreviated quantum field theory.  The correspondences thus
established should make clear both the motivation for the particular
rules of the CTC mechanics and the appropriate generalization to more
realistic systems.

In the absence of even an existence proof, there is certainly no
``realistic'' candidate for a spacetime.  Hence, I choose one of a
variety of possible generic structures.$^{[9]}$  Begin with a flat
spacetime.  Identify a compact region of space at time $t = 0$ with a
compact region of space at $t = T$ such that world lines entering the
identified region from $t \lsim 0$ connect smoothly and immediately to
the identified point at $t = T$ and continue with $t > T;$ world lines
entering the identified region from $t \lsim T$ connect smoothly and
immediately to the identified point at $t = 0$ and continue with $t >
0$.  If the identified regions are the same, then the obvious
point-by-point identification leaves the spacetime flat except for
conical singularities.  This construction is illustrated in the
accompanying figure for one-plus-one dimensions.  There are four classes
of straight world lines: ~1) lines uneffected by the identification
(these may either miss the time machine altogether or pass between
the identified regions so
quickly as to be unaltered); ~2) lines that jump from $t = 0$ to $t = T$
without ever passing in between; ~3) lines that first encounter the time
machine at $t = T$ and wind around some number of times before exiting
to $t \rightarrow \infty$; and ~4) CTC's that have no existence for $t <
0$ or $t > T$.

The first drastic simplification is to reduce the above described space
to two points.  One point, $z_1$, is outside the time machine, with a
single continuous time history on the interval $- \infty < t < \infty$.  The
other point, $z_2$, corresponds to the identified region of space.  It
has  two disjoint segments to its history.  One is a continuous line
from $t = - \infty$ to $t = \infty$ excluding $0 < t < T$.  The other is the
CTC formed by identifying the spacetime point at $t = T$ with the point
at $t = 0$.

The second simplification concerns the nature of the quantum
field defined on the space.  In second quantization, bosons are
described by an oscillator degree of freedom at each point in space.
Even for a single point, this corresponds to an infinite dimensional
Hilbert space.  To make matters simpler I choose to consider a single
fermion field.  At each spatial point a single fermionic degree of
freedom corresponds to a two-dimensional Hilbert space, occupied or
unoccupied in the language of second quantized fermions or,
equivalently, spin up or down.  With only two sites, the total Hilbert
space has only four dimensions.

To make the connection of the thus defined two spin system to quantum
field theory and to make explicit the meaning of a sum over histories
for the spins, I briefly review fermion path integrals,$^{[10]}$ in the
absence of CTC's.

Fermions can be described by a Grassman number valued field
$\psi(z,t)$.  In the present case the positions $z$ take only two
values, $z_1$ and $z_2$.  So there really are only two Grassman (fully
anti-commuting) coordinates:
$$\eqalign{x(t) &\equiv \psi (z_1,t)\cr
y(t) &\equiv \psi (z_2,t)\,\, .\cr}$$
We will also need the independent, conjugate coordinates $\bar x$ and
$\bar y$.  An alternative description is given by the set of operators
$a, a^{\dag}, b$, and $b^{\dag}$, corresponding respectively to $x, \bar
x, y,$ and $\bar y$.  These anticommute with themselves and each other
except for the canonical anticommutators
$$\eqalign{[a, a^{\dag}] &= 1\cr
[b, b^{\dag}] &= 1 \,\, . \cr}$$
The most general wave function is
$$	\phi(x,y) = c_0 + c_1 x + c_2 y + c_3 xy\,\, ,$$
with complex coefficients $c_i$, defining a vector space equivalent to
that of two half-integer spins.

The most general Hamiltonian, $H(a, a^{\dag}, b, b^{\dag})$, contains
terms zero-through-fourth degree in the operators and is parametrized by
sixteen real numbers, corresponding to the general four-by-four
Hermetian matrix Hamiltonian for the two-spin system.

The path integral that generates quantum amplitudes and, in particular,
the evolution kernel, is an integral over the Grassman fields
$\psi(z,t)$ and $\bar \psi(z,t)$ or, in the present case, Grassman paths
$x(t), \bar x(t), y(t)$, and $\bar y(t)$, with suitable boundary
conditions,$^{[10]}$ of $exp (iS)$,  where the action
$$	S = \int dt [- \bar x i\partial_t x - \bar y i\partial_t y
- H(x, \bar x, y, \bar y)].$$

The field theoretic perturbative expansion expresses all integrals in
terms of Gaussians.  Having only two sites reduces the problem
drastically: the model is soluble in as much as one can diagonalize a
four-by-four matrix.

For illustrative purposes it will suffice to consider only a few
particular examples within the class of possible Hamiltonians.  In
particular, I will analyze special cases that can be diagonalized by
inspection of
$$	\eqalign{H &= \omega_0 (a^{\dag} a + b^{\dag} b) + \omega_1
(a^{\dag} b + b^{\dag} a)\cr
& + \gamma_1 a^{\dag} a (b + b^{\dag}) + \gamma_2 (a + a^{\dag})
b^{\dag} b\cr
&+ \lambda a^{\dag} a b^{\dag} b\,\, ,\cr}\eqno (1)$$
where $\omega_{0,1}, \gamma_{1,2}$, and $\lambda$ are real parameters.
Define a basis set of states for the two spins:
$$	\{|\uparrow\uparrow \rangle, |\uparrow\downarrow \rangle,
|\downarrow\uparrow \rangle, |\downarrow\downarrow \rangle\} \eqno (2)$$
by
$$	\eqalign{a | \downarrow i \rangle &= 0 \qquad \qquad a^{\dag}
|\downarrow i \rangle = | \uparrow i \rangle\cr
b | i \downarrow \rangle &= 0 \qquad \qquad b^{\dag} | i \downarrow
\rangle = |i
\uparrow \rangle \,\, , \cr}$$
etc., where $i = \uparrow$ or $\downarrow$.  In this basis, the
Hamiltonian in eq. (1) is represented by
$$	H = \pmatrix{2\omega_0 + \lambda & \gamma_1 & \gamma_2 & 0\cr
\gamma_1 & \omega_0 & \omega_1 & 0\cr
\gamma_2 & \omega_1 & \omega_0 & 0\cr
0 & 0 & 0 & 0\cr}\,\, . \eqno (3)$$
In the language of second quantized fields, $\omega_0$
corresponds to the particle's mass.  The $\omega_1$ term is analogous to
the kinetic or spatial hopping term.  $\gamma_{1,2}$ are Yukawa-like
couplings (not normally allowed for purely fermionic interactions).
$\lambda$ is the unique analog in this system of a four-fermion
coupling.  The $\omega$'s multiply linear terms in the field equations
or Euler-Lagrange equations of motion for the Grassman paths, while
$\gamma_{1,2}$ and $\lambda$ introduce nonlinearities.

In the presence of CTC's, there is no foliation of the spacetime (unique
time ordering of spacelike surfaces) and hence no Hamiltonian evolution
or Schr\"odinger equation.  Nevertheless, both versions of quantum
mechanics discussed below can be compactly characterized in terms of the
action of the Hamiltonian $H$ and, in particular, the unitary operator
$$	U(t', t) \equiv e^{-iH(t' - t)} \,\, . \eqno (4)$$

\noindent {\bf Path Integral for Pure Initial States}

The Feynman path integral offers a geometrically appealing
generalization of quantum mechanics for a CTC -- containing
spacetime that does not admit
a Schr\"odinger equation.  One sums coherently over all field histories
or paths defined over the spacetime, weighted by ${\it exp}$ ($i$
Action).\foot{When considering particle dynamics, one must not confuse
an essentially first quantized description in which particle world lines
may wind any number of times around the CTC region from a second
quantized description in which the fields (describing any number of
particles) are single valued on the spacetime.  The model considered
here is of the latter type.}  Correlations between observables are
determined from the amplitudes obtained by projecting the paths onto the
subsets that satisfy the possible observation outcomes.  In principle,
this defines all correlations, including those for measurements made
during the CTC epoch.  However, I will only consider observations made
either before or after a compact CTC region.  While the phenomena within the
CTC epoch itself may offend our sensibilities, to dismiss their
possibility on that basis may be premature.

There are three logical steps to the full construction of the
mechanics.  ~1) Consider an amplitude defined by initial and final
states (before and after the CTC's) as boundary conditions on the path
integral.  We will find that systems with non-linear equations of motion
have a non-unitary initial-to-final map, as defined by the path
integral.  ~2) To re-establish a probability
interpretation of the (non-unitary) amplitudes -- squared,
we renormalize the sum of
the probabilities for all possible outcomes for each individual initial
condition to one.   Thus we determine the probability for any particular
outcome relative to any particular initial condition.  However, this
renormalization depends on the initial state.  Hence the linearity of
quantum mechanics, i.e., the superposition principle, is lost.  And  ~3)
to maintain a consistent probability calculus in light of step 2, the
path integral must include all paths extending to the future of the
compact CTC region -- even in computing correlations of only
observations completed before the CTC's.  It is this last step that has
the most bizarre consequences:  Pre-CTC quantum mechanics is non-linear,
non-unitary, and acausal, even in the absence of visits, real or
virtual, from time travelers.

So, to begin I evaluate the path integral connecting states at $t = -
\epsilon \equiv 0^-$ to states at $t = T + \epsilon$, with $\epsilon
\rightarrow 0^+$.  If $z_2$ is the spatial location of the time machine,
then the field at $z_2$ satisfies $\psi (z_2, 0^-) = y(0^-) = \psi (z_2,
T^+) = y(T^+)$, by construction.  Hence, we need only evaluate how
various values of $\psi (z_1, T^-) = x (0^-)$ connect to $\psi (z_1,
T^+) = x (T^+)$.  In fact, in this path integral discussion, I will
henceforth suppress the dependence on the field at $z_2$ outside the
interval $0 \leq t \leq T$.  Explicit restoration of that dependence
would be totally straightforward: using a product basis, such as
provided by eq. (2), one simply multiplies the emerging $x(T^+)$ by the
incident $y(T^-)$ for each basis state and then adds the components.

A simple strategy for explicit evaluation of the functional integral is
to slice up the spacetime, do the integral over the slices with general
boundary conditions, match the slices, and integrate over the boundary
conditions of the interfaces to reconstruct the full integral.  The
integration over Grassman positions $x$ and $y$ at a given $t$ is really
a sum over a complete set of states.  Hence, for convenience we can use
the spin basis defined by eq. (2) to specify boundary conditions.
And finally, the path integral over a finite time interval containing no
CTC's is given by the action of $U$ defined in eq. (4).  Putting
all this together and defining the operator $X$ as taking states of $z_1$
at $t = 0^-$ to $z_1$ at $T^+$, one deduces the corresponding path
integral to equal
$$	\langle j |X|i\rangle = \sum_k \langle jk|U (T,0) | ik\rangle
\,\, , \eqno (5)$$
where $i,j,k = \uparrow$ or $\downarrow$.

It is a simple exercise to see that the linear field theory defined by
$\omega_{0,1} \not= 0$ and $\gamma_{1,2} = \lambda = 0$ in eq. (3)
yields a unitary $X$.

Note that ``unitarity" should include the possibility that $X^{\dag} X$ is
proportional to and not only equal to the identity because in that case
a single, state-independent factor, which could be absorbed into the
functional measure, can restore literal unitarity without altering any
observable correlations.  A proof that all such linear theories produce
unitary $X$'s goes as follows.  The most general Hamiltonian with linear
equations of motion is
$$	\eqalign{H &= r_1 + c_1 a + c^*_1 a^{\dag} + c_2 b + c^*_2 b^{\dag}\cr
&+ r_2 a^{\dag} a + r_3 b^{\dag} b + c_3 ab + c^*_3 b^{\dag} a^{\dag}\cr
&+ c_4 a^{\dag} b + c^*_4 b^{\dag} a\,\, ,\cr}$$
where $r_i$ and $c_i$ are real and complex coefficients.  Without loss
of generality, we can set $c_1 = c_2 = 0$; this corresponds to shifting
the operators by Grassman $c$-numbers or mixing {\it a}
 with $a^{\dag}$ and mixing $b$ with
$b^{\dag}$ to define a new basis for each individual spin.  Written in
four-by-four matrix form, the resulting $H$ is block-diagonal, i.e., $H
= h \otimes h'$, where $h$ acts in the $(\uparrow \uparrow, \downarrow
\downarrow)$ space and $h'$ acts in the $(\uparrow \downarrow,
\downarrow \uparrow)$ space.  Also, $tr ~h = tr ~h'$, which is a further
crucial consequence of the linearity.  Hence, the unitary evolution in
the four-dimensional space is of the form $U = u \otimes u'$, where $u$
and $u'$ are unitary two-by-two matrices on the above mentioned
subspaces, and $\det u = \det u'$.  It is this equality of determinants
that ensures the unitarity of $X$ upon performing the partial trace
prescription of eq. (5). (To do the traces explicitly, one can
write $u$ and $u^{\prime}$ as the appropriate sums of the Pauli and unit
matrices and implement the equality of determinants.)

The nonlinear couplings, e.g., $\gamma_{1,2}$ and $\lambda$, produce a
non-unitary $X$.  A very simple example of a non-unitary $X$ follows
from $\lambda \not= 0$ but $\omega_{0,1} = \gamma_{1,2} = 0$.  I will
use this for illustrative purposes later with a maximally non-unitary
value $\lambda T = \pi + 2 \epsilon \approx \pi$; in this case
$$	\eqalign{X &= \pmatrix{1/2 (1 + e^{-i\lambda T}) & 0\cr
0 & 1\cr}\cr
&\approx \pmatrix{i\epsilon & 0\cr
0 & 1\cr}\,\, . \cr} \eqno (6)$$

In general, if a non-unitary evolution operator $X$ maps an initial
state $\Psi$ onto a final state $X\Psi$, then we must renormalize the
final state to become $X\Psi/(\Psi^{\dag} X^{\dag} X \Psi)^{1/2}$.  For
non-unitary $X$, the evolution is, consequently, non-linear in $\Psi$.
Given that $\Psi$ is a pure quantum state, this final state is also pure
(as one may confirm by computing its density matrix).

The implementation of initial-state-dependent renormalizations to
generate a probability interpretation as outlined in step (2) above is
straightforward.  What may not be so clear is the necessity of
Hartle's$^{[8]}$ assertion that even correlations before the CTC's are
influenced by the non-unitarity.  For example, in the system at hand,
why couldn't the correlation of the spins at $t_2$ with their values at
$t_1$ for $t_{1,2} < 0$ (i.e., before the CTC) be computed using the path
integral from $t_1$ to $t_2$ with the appropriate states as boundary
conditions?  (This calculation would give the results of conventional
quantum mechanics.)  Instead, we are instructed to evaluate a path
integral that traverses the CTC region.  The most compelling argument
for this procedure is not, I believe, given explicitly by Hartle but is
implicit in the desire to produce a consistent probability calculus.  In
particular, if it is possible to make observations after the CTC epoch,
then it is reasonable to require that observations made before the CTC's
are consistent with a larger set of observations that includes
measurement after as well as before.  I offer the following example to
illustrate the construction and its consequences.

It is sufficient to consider a single spin, up or down, whose path
integral from $t = 0$ to $t = T$ is given by a non-unitary $X$, e.g.,
that of eq. (6).  In this example I take the dynamics of the single spin
before $t = 0$ and after $t = T$ to be trivial, i.e., given by the identity
operator; so up and down are degenerate outside $0 \leq t \leq T$.
Consider times $t_1 \leq t_2
\leq 0$ and $t_3 > T$.  Given the initial condition that the state at
$t_1$ is
$$	| + \rangle \equiv {1\over \sqrt{2}}~ (| \uparrow \rangle
+ | \downarrow \rangle)
\,\, ,$$
what are the probabilities for $\uparrow$ and $\downarrow$ at $t_2$?
Without $X$, they would both be $1/2$.  However, if we consider
continuing paths to $ t_3$ and measuring $\uparrow$ versus $\downarrow$
at that time, we would find the probabilities ${\cal P} (i,j,k)$, where
$i,j,k$ are the spins at $t_1, t_2, t_3$ respectively,
$$	\eqalign{{\cal P} (+, \uparrow, \uparrow) &= \epsilon^2/(1 +
\epsilon^2)\cr
{\cal P} (+, \uparrow, \downarrow) &= 0\cr
{\cal P} (+, \downarrow, \uparrow) &= 0\cr
{\cal P} (+, \downarrow, \downarrow) &= 1/(1 + \epsilon^2) \,\, . \cr}$$
These follow from evaluating the path integral with boundary conditions
$i$ and $k$ and projecting onto $j$; squaring; and then renormalizing
the sum to one.  ($X$ would be unitary were $\epsilon = 1$.)  We now
require that the probability of spin $j$ at $t_2$ given spin $i$ at
$t_1$ is given by
$$	{\cal P} (i,j) = \sum_k {\cal P} (i,j,k) \,\, . $$
Therefore,
$$	{\cal P} (+, \uparrow) = \epsilon^2/(1 + \epsilon^2)$$
$$	{\cal P} (+, \downarrow) = 1/(1 + \epsilon^2) \,\, .$$
This is, perhaps, disconcerting, particularly as we could have chosen
$t_2$ close to, or even equal to, $t_1$.  However, it is not only
logically consistent but also in keeping with the original requirement
of only allowing consistent world histories.$^{[11]}$  The situation is
simplest to describe in words for the extreme case of $\epsilon = 0$.
For $\epsilon = 0$, trajectories on which the state $\uparrow$ enters
the time machine interfere destructively with each other.
Mathematically, we can always decompose the state $+$ into a
superposition of $\uparrow$ and $\downarrow$.  However, if we actually
perform
an $\uparrow$ or $\downarrow$ {\it measurement} (i.e., a physical
process) on the state $+$, the full
world histories for which the outcome of the measurement is $\uparrow$
self-destruct and together carry vanishing probability.

We can now resolve the apparent discontinuity at $\epsilon = 0$.  For
$\epsilon
\not= 0$, if a state approaches $t = 0$ as pure $\uparrow$, it will
leave $t = T$ as pure $\uparrow$ with probability one.  Yet for
$\epsilon = 0$,
there is no amplitude for an $\uparrow$ component at $t = T$,
irrespective of initial state.  However, from the previous discussion, we
learned that as $\epsilon \rightarrow 0$, it becomes harder and harder to
prepare a pure $\uparrow$ state at $t = 0$.  Hence, the physics of
$\epsilon = 0$ connects smoothly to the behavior as $\epsilon \rightarrow
0$.

Clearly, with this sort of mechanics, suitable experiments before the CTC
epoch could determine that CTC's will be in their future.  Presumably,
with a time machine of finite extent within a continuous space, the
strength of this effect would be proportional to the fraction of the
experiment's future light cone that intersects the time machine.  Hence,
the observed linearity of quantum mechanics, although verified in
experiments of enormous precision,
does not in
practice tell us much about future CTC's, except that they are not in
our immediate future.

\noindent {\bf Density Matrix Mechanics}

Deutsch proposed$^{[2]}$ a profoundly different mechanics that likewise
reduces to ordinary quantum mechanics in the absence of CTC's.  One
considers a density matrix $\rho$ for the entire system that evolves in
the standard fashion,
$$	i\partial_t \rho = [H,\rho]$$
or
$$	\rho (t') = U\rho (t) U^{\dag}\,\, ,$$
where $U$ is given in eq. (4).  So the dynamics is linear in
$\rho$, and Hermeticity of $H$ ensures that $\rho$ remains normalized to
$tr~ \rho = 1$.  The CTC's enter in boundary and consistency conditions
on $\rho$.  If the degrees of freedom are in two disjoint sets labeled 1
and 2 (in our simple model they are just the two spins) and set 2 refers
to the CTC's, then we require
$$	tr_1 \rho (0^+) = tr_1 \rho (T^-) \,\,, \eqno (7)$$
where $tr_i, ~i = 1$ or $2$, is the partial trace over the labeled
subset.  I will return for successive clarifications on the handling of
initial conditions, but, roughly speaking, we will do something like,
for a given incoming $\rho(0^-)$,
$$	tr_2 \rho (0^-) = tr_2 \rho (0^+) \eqno (8)$$
$$	tr_1 \rho (0^-) = tr_1 \rho (T^+) \,\, . \eqno (9)$$
$tr_1 \rho (0^+)$ is not set by initial conditions but rather by the
consistency of information around the CTC, implemented by eq. (7),
which gives as many linear constraints as there are free parameters in
$tr_1 \rho (0^+)$.  As determined by the evolution equations, the
dynamical output is, roughly,
$$	tr_2 \rho (T^+) = tr_2 \rho (T^-)\,\, . \eqno (10)$$
The evolution of $\rho$ outside a CTC epoch is identical to ordinary
quantum mechanics, and CTC's do not generically impose any {\it a
posteriori}
constraint on physics in their past$^{[2]}$.  However, if the density
sub-matrix
$tr_1 \rho (T^-)$ is not pure,\foot{A pure density matrix $\rho$
satisfies $tr \rho^2 = 1$ and can be written $\rho = | \Psi
\rangle\langle \Psi |$
where $|\Psi \rangle$ is some pure state.} the matching provided by eq.
(7) does not require the matching of states, amplitudes, or paths around
the CTC.  In general, pure states will evolve into mixed states, as
illustrated in examples that follow.

First consider an initial density matrix $\rho_1 (0) \equiv tr_2 \rho
(0^-)$ that is pure, and ignore the 2 variable at point $z_2$ for $t =
0^-$.  What is $\rho_1 (T) \equiv tr_2 \rho (T^-)$ using the time
evolution and eq. (7) for various $H$'s or $U$'s?

To proceed, one needs a simple theorem:  given a system composed of two
subsystems, 1 and 2, the most general total density matrix $\rho$
compatible with $tr_2 \rho$ being pure is $(tr_2 \rho) \otimes \rho_2$
where $\rho_2$ is the most general density matrix for subsystem 2.

As a first example, consider $H$ and, therefore, $U$ which are diagonal
in the basis defined by eq. (2).  The necessary calculation is a
simple exercise in four-by-four matrix multiplication, taking partial
traces, and solving linear equations.  For this particular class of
$U$'s, one finds that the sub-matrix $tr_1 \rho$ is not fully determined
by the matching condition, eq. (7).  (The implied linear conditions
are not all linearly independent.)  Nevertheless, if the spin at $z_1$ is
initially in the pure $\uparrow$ state, it necessarily emerges at $T$ as pure
$\uparrow$, and pure initial $\downarrow$ emerges as pure $\downarrow$.
For other pure initial states, there exists a solution of eq. (7)
for which they emerge at $T$ unchanged.  Hence, quantum coherence is not
necessarily destroyed.  However, there exist solutions of eq. (7)
for which initial non-trivial linear combinations of $\uparrow$ and
$\downarrow$ emerge as mixed states, i.e., $tr_2 \rho (T^-)$ is not
pure.  Deutsch proposes to resolve such ambiguities by maximizing the
entropy.  One could just as soon choose a unique solution by minimizing
the entropy.  Rather than either of these, I would stress that this
indeterminacy arises for a set of measure zero in the parameter space
of generic $H$'s.  The neighborhood of that set will either provide a
unique limiting behavior, or the set will be a separatrix of different
qualitative behaviors.  In either case, the dynamics would provide a
resolution or interpretation of the ambiguity.

With a richer class of interactions, quantum decoherence is a necessary
consequence of the proposed density matrix mechanics.  A simple explicit
example is provided by
$$	U = \pmatrix{1 & 0 & 0 & 0\cr
0 & 1/\sqrt{2} & 1/\sqrt{2} & 0 \cr
0 & -1/\sqrt{2} & 1/\sqrt{2} & 0 \cr
0 & 0 & 0 &  1\cr}\,\, , \eqno (11)$$
which could arise, say, from $H$ of eq. (3) with $\omega_0 =
\gamma_{1,2} = \lambda = 0$ and $\omega_1 T = \pi/4$, an example with
linear equations of motion and unitary evolution according to the path
integral approach.  Again, the pure states $\uparrow$ and $\downarrow$
evolve into themselves.  However, if the spin at $z_1$ is initially in the
pure $+$ state $(| \uparrow \rangle + | \downarrow \rangle)/\sqrt{2}$, its
final
density matrix is
$$	\rho_1 (T^+) = tr_2 \rho (T^-) = \pmatrix{1/2 & 1/2\sqrt{2}\cr
1/2\sqrt{2} & 1/2\cr} \,\, ,$$
which is decidedly mixed.

Turn now to the case of an input density matrix that is mixed, which
could arise either because of the initial correlations of spins 1 and 2
in a pure state before the CTC or because the total state is mixed.  It
is here that I will argue for a departure from Deutsch's calculus.  To
make the comparison clearest, I begin with a system identical to that
analyzed in Reference 2.  Deutsch describes the system as a 1/2 integer
quantum spin carried along a prescribed classical trajectory that
intersects itself once as it loops once around a time machine.  Hence,
the Hilbert space outside the CTC epoch consists of a single spin.  The
evolution of the two spins upon contact within the CTC epoch is given by
some general unitary transformation, which can be thought of as the
product of the $U$ matrix discussed above with the unitary matrix
$$	V = \pmatrix{1 & 0 & 0 & 0\cr
0 & 0 & 1 & 0\cr
0 & 1 & 0 & 0 \cr
0 & 0 & 0 & 1 \cr}\,\, ,$$
which interchanges 1 and 2.  This label switching arises because of
Deutsch's parametrizing the classical trajectory by its proper time and
then describing the spins as ``younger'' and ``older.''  The CTC allows
the outgoing younger spin to enter as the older one.
In contrast, my labels, 1 and 2, preserve their meaning around the CTC.

The initial conditions are completely characterized by the $t = 0$
density matrix for spin number 1, $~\rho_1 (0)$.  If $\rho_1(0)$ is mixed, then
the most general density matrix $\rho (0)$ for the 1,2 system such that
$tr_2 \rho (0) = \rho_1 (0)$ is no longer of the product form, $\rho_1
(0) \otimes \rho_2 (0)$, and, in fact, it has many more free
parameters.  Deutsch argues that we should nevertheless assume a product
form and proceed as before.  (Following this prescription of assuming a
product form, for generic values of the parameters,
$tr_1 \rho (0)$ will be determined by the consistency condition,
eq. (7), and the output $\rho_1 (T) = tr_2 \rho (T)$ will be unique
and, in fact, mixed.)  The justification of the assumed restriction on
$\rho (0)$ to the product form is the rhetorical, ``How could spin 2 be
correlated with spin 1 at $t = 0$ when they had not yet had a chance to
interact?''

Deutsch's construction yields a logically self-consistent dynamics.
However, I will now argue in favor of an alternate approach that has the
virtue of restoring a traditional ensemble and probability
interpretation to the density matrix that was otherwise lost.  -- The
rhetorical retort is that at $t = 0$ spin 2 had interacted with spin 1
in its past, even if those interactions weren't in spin 1's past.

In normal quantum theory, the density matrix can be viewed as describing
a statistical ensemble of pure states.  One can combine ensembles by
adding together their $\rho$'s, weighted by relative probabilities.
Also, one can decompose an ensemble into subsets.  This corresponds to
dividing $\rho$ into pieces, a process limited only by the restriction
that the component $\rho$'s are themselves legitimate density matrices.
\foot{A density matrix is Hermetian, has unit trace, and has
eigenvalues (and hence, diagonal elements) between 0 and 1.}
Such composition or decomposition of ensembles commutes with the time
evolution because the dynamical equations are linear in $\rho$.  Since
the proposed time evolution of $\rho$ is the conventional, linear one,
can this ensemble and probability interpretation be preserved?

A natural decomposition of a mixed $\rho$ is to consider it as a
statistical, incoherent ensemble of its eigenstates, each with a
probability given by its respective eigenvalue.  Each individual
eigenstate, of course, defines its own pure $\rho$, and we disussed
earlier the density matrix mechanics propagation of a pure initial
$\rho$ across a CTC epoch.  In that case, the absence of correlation
between 1 and 2 was a mathematical necessity, not an additional
assumption.  If we consider a mixed $\rho_1 (0)$ as an ensemble of its
pure eigenstates, propagate each of them separately to $t = T$ to obtain
a (generically) unique $\rho_1 (T)$, and then add those together
weighted by the initial probabilities,  we get a final answer which is
itself a solution of the total evolution and matching conditions.  It
differs from Deutsch's answer because the total $\rho (0)$ is not of the
pure product form $\rho_1 (0) \otimes \rho_2 (0)$.  Instead, it is a
particular sum of products.  Certainly not the most general $\rho (0)$
satisfying $tr_2 \rho (0) = \rho_1 (0)$, its value is, nevertheless,
determined by $\rho_1 (0)$.
It is this $\rho (0)$, and not the pure product form, that allows $\rho$
to be interpreted in terms of an ensemble with additive probabilities.
Deutsch's pure product algorithm determines a $\rho$ that encodes the
predictions for all possible observations, but the simple linear
mathematics representing the physical acts of combining and dividing
ensembles is lost.

Return, now, to the system originally defined in this paper, in which
the spin at $z_2$ has an existence outside the CTC epoch, and consider
pure initial states of the two-spin system such that $\rho_1 (0^-)$ is
mixed.  Were we to treat $\rho_1 (0^-)$ as an initial condition as
described in the preceding paragraphs, we would by construction be
destroying the coherence of the initial total state.  It would be
equivalent to replacing the initially pure total $\rho (0)$ by $tr_2
\rho \otimes tr_1 \rho (0)$, which is, typically, mixed.  While this may
be a tenable proposal for CTC physics, I will pursue here the
construction of a dynamics that does not discard information
willy--nilly but only loses it when absolutely necessary.

A natural way of keeping track of at least some of the pure initial
state 1--2 correlations upon entering the CTC epoch is given by the
Schmidt decomposition, which provides the following.  A pure state
$|\Psi \rangle$ of a system composed of subsystems 1 and 2 can be written
$$| \Psi\rangle = \sum_a ~ c_a |a\rangle_1 \otimes |a\rangle_2 \,\, ,$$
where the set of states $|a\rangle_2$ form an orthonormal basis for the
subspace 2 (chosen to be the smaller of spaces 1 and 2, if they have
different dimensionality); the set of $|a\rangle_1$ are an equal number of
orthonormal states in space 1; and $c_a$ are complex coefficients.  (The
proof by construction begins by choosing the $|a\rangle_2$ as the eigenvectors
of $tr_1 |\Psi\rangle \langle\Psi|$.  One then writes $|\Psi \rangle$ in
the most general possible form using the $|a\rangle_2$'s and some basis
on subspace 1.  Finally, one requires that $tr_1
|\Psi\rangle\langle\Psi|$ has the initially assumed form.)

Apply this decomposition to the problem at hand.  If $\rho (0^-) =
|\Psi\rangle\langle \Psi|$, then
$$tr_2 \rho(0^-) = \sum_a |c_a|^2|a\rangle_1{_1\langle}a|\,\,.$$
This density matrix for the spin at $z_1$ is not in general pure.
However, it is here written in diagonal form which can be interpreted,
as argued previously, as representing an ensemble of pure states,
labeled by $a$ and with probabilities $|c_a|^2$.  Recalling the Schmidt
decomposition of $|\Psi\rangle$, we know that each $|a\rangle_1$ comes with its
particular $|a\rangle_2$.  Each pure sub-ensemble $|a\rangle_1
{_1\langle}a|$ at $ t = 0$
is mapped by the dynamical evolution to an (in general, mixed) state
specified by some $\rho^a_1 (T^+)$.  So we can take the final total
density matrix to be
$$\rho(T^+) = \sum_a |c_a|^2~\rho^a_1 (T^+) \otimes
|a\rangle_2{_2\langle}a|\,\,.\eqno(12)$$

There is an important lesson implicit in eq. (12), our prescription for
propagating general, pure initial states.  While it was always clear
that the matching of $tr_1 \rho$ at $t = 0^+$ and $T^-$ was a potential
source of incoherence (because we deal only with information encoded
locally in density matrices), the matching of $tr_1 \rho$ at $t = 0^-$
and $T^+$ likewise destroys coherence.  Those initial correlations
between the two spins that are encoded as relative phase information in
the language of wave functions or in certain off-diagonal elements of the
total density matrix are simply lost when crossing the CTC epoch if we
insist on using only density matrices, partial traces and differential
time evolution.  A simple example is provided by the pure initial state
$$|\Psi\rangle = (|\uparrow \downarrow \rangle - | \downarrow \uparrow
\rangle)/\sqrt 2$$
or
$$	\rho(0^-) = {1 \over 2}\pmatrix{0&0&0&0\cr
0&1&-1&0\cr
0&-1&1&0\cr
0&0&0&0\cr}\,\,.$$
Even under the trivial dynamics of $U = 1$ (for which nothing happens to
any state), the output $\rho (T^+)$ is mixed.  Were we to simply
propagate the two partial traces, since each individually is maximally
random, the total $\rho(T^+)$ would be maximally random.  In contrast, the
construction that lead to eq. (12) allows us to preserve the fact that
$\uparrow$ comes with $\downarrow$ and {\it vice versa},
$$\rho (T^+) = {1 \over 2} \pmatrix{0&0&0&0\cr
0&1&0&0\cr
0&0&1&0\cr
0&0&0&0}\,\, ,$$
but we have lost the phase information that distinguishes $|\Psi\rangle$ from
\hfill\break
$(|\uparrow\downarrow\rangle + |\downarrow\uparrow\rangle)/\sqrt 2$.

Mixed initial states can be decomposed as a sum of pure ensembles, given
by the $\rho (0^-)$ eigenvectors and weighted by their eigenvalues.
Each pure state can be propagated as described above and the ensemble
reassembled, thus defining a density matrix mechanics which minimizes
information loss and preserves the ensemble interpretation.

\noindent{\bf Mixed States in the Path Integral Approach}

The evolution of pure states within the path integral approach was
specified in our earlier discussion.  In particular, if non-unitarity is
confined to path integrals over a particular epoch, $0 \leq t \leq T$,
then an initial state specified at some $t_0 < 0$ can serve as the
initial boundary condition on the path integral.  Unitarity within the
$t < 0$ epoch ensures that starting the path integral from a yet earlier
time and keeping only those paths consistent with the specification at $t_0$
would
yield identical predictions.  In contrast, the specification of a mixed
initial state and, more generally, many considerations regarding information
flow must be altered if the dynamics includes a non-unitary epoch.

If $X$ is the non-unitary operator that maps states at $t = 0$ to states
at $t = T$, then one might imagine that a density matrix $\rho (0)$
characterizing a mixed state at $t = 0$ is mapped to $t = T$ according
to
$$	\rho (T) = X \rho (0) X^{\dag}/tr X\rho (0) X^{\dag}\,\, .$$
However, this is untenable.  It suffices to consider a two-state system
as an example, with $X$ given by eq. (6).  If the system at $t = 0$
has equal probabilities to be $\uparrow$ and to be $\downarrow$, then
the system evolves into the statistically identical mixed state
at $t = T$ because $X$ maps
$\uparrow$ onto $\uparrow$ and $\downarrow$ onto $\downarrow$.  However,
if the system is equally likely to be $+$ or $-$ (with $|\pm\rangle =
(|\uparrow \rangle \pm |\downarrow \rangle)/\sqrt{2})$ at $t = 0$, then the
system
at $t = T$ is nearly pure  $\downarrow$.  These two ensembles have the
same density matrix at $t = 0$, but differ at $t = T$.  In contrast to
normal quantum mechanics, knowing the density matrix at one time (and,
therefore, the mean value of all observables at that time) does not
determine the density matrix at a future time, even if the evolution of
all pure states is known.  Hence, a statistical ensemble is not
completely characterized by its density matrix but rather must be
specified in terms of probabilities for particular (not necessarily
orthogonal) states.

In light of these considerations, with non-unitary evolution the
traditional entropy
$$	\sigma \equiv - tr \rho \log \rho$$
cannot be an adequate measure of disorder or lack of information because
$\rho$ does not characterize the statistical state of the system.
Rather than attempt to propose an adequate alternative, I comment here
on two antithetical  perspectives, again posed in the explicit context
of the example provided by eq. (6).

The non-unitary $X$ of eq. (6) maps nearly all pure states onto
states that are nearly pure $\downarrow$.  Does this imply that the
information in a typical situation increases in going from $t = 0$ to $t
= T$?  The answer depends on what is a suitable measure (characterizing
uniform distribution of probability) in the space of states.  Were the
states actually half-integer intrinsic angular momenta, then the Haar
measure on the rotation group would provide a natural measure in the
spin space.  The situation is not so clear if we are dealing with
occupied and unoccupied modes of a Fermi field.
``Totally random" does not necessarily mean equal probabilities for
occupied and unoccupied.  ``Randomness" in a physical system must be
defined relative to some physical process or situation, e.g., high
temperature.
Furthermore, given
any measure at $t = 0$, $X$ itself defines a new measure at $t = T$
(dynamically transporting the $t = 0$ measure) such that $X$ is volume
preserving or, equivalently, information preserving.

On the other hand, one may be concerned that $X$ typically destroys
information in the following sense:  if a message is encoded in a
particular state (or sequence of states) before $t = 0$, what arrives at
$t = T$ is nearly independent of the information encoded.  What started
out as orthogonal alternatives are now nearly identical, and unambiguous
reading of the message seems hopeless.  However, the key word here is
``nearly,'' because if CTC's are, indeed, possible, the ``arbitrarily
technologically advanced civilization''$^{[11]}$  could construct a
100\% efficient message decoder (essentially containing a version of
$X^{-1}$).  Hence, $X$ can again be viewed as information preserving.

\noindent {\bf Conclusions}

The initial motivation for studying possible dynamics in the presence of
CTC's was to search for irresolvable paradoxes which might preclude the
existence of any consistent dynamics and, hence, of CTC's themselves.
Yet, rather than identifying some fatal, irresolvable contradiction, we
encountered a proliferation of alterative dynamical schemes, each
reducing to ordinary quantum mechanics in the absence of CTC's.

The method used here was to consider a very simple system that could be
solved explicitly but was rich enough to provide examples of interesting
phenomena and had a structure that would make the generalization to more
sophisticated systems fairly self-evident.  The two main directions
explored here had been suggested previously.  One uses a path integral
and has coherent states without a Schr\"odinger equation.  The other uses
density matrices and has Hamiltonian evolution without coherent states.
In each case, the CTC's produce bizarre (and different) phenomena, even
outside the CTC epoch.

There are certainly potentially interesting questions that one cannot
address within models as simple as those analyzed here.  In a spacetime
with a non-uniform metric there can be particle production.  What now
happens to pure initial states, e.g., the ground state, in the various
schemes?  What sort of conservation laws survive in the presence of
CTC's?

The dynamical systems studied here may provide helpful examples regarding
issues of complexity.  Many discussions of complexity in quantum
mechanics focus on features that can emerge only in large dimensional
vector spaces, such as the statistics of level spacings.  However, in
the above discussion we have examples of four-by-four matrix
Hamiltonians that fall into qualitatively different dynamical classes.
What, in a small matrix, can distinguish simple from complex behavior?
In the path integral investigation, the aspect that distinguished
unitary from non-unitary (with its decidedly eerie consequences) was
linearity of the  equations of motion.  In the density matrix
formulation, incoherence arose because of a clash between the eigenstates
of the Hamiltonian and the basis for the Hilbert space that was
natural relative to the spacetime.  To handle the CTC, which treated
spatial points $z_1$ and $z_2$ differently, it was convenient to use a
Hilbert space basis that factored into products of the spin at $z$, with the
spin
at $z_2$.  If we are allowed to consider (as in this example) the
spacetime and the concept of locality as specified independently of the
dynamics, then the extent to which the dynamics respects that locality
is a measure of its simplicity.  If spacetime is itself part of the
dynamics (as it must ultimately be), then the only useful analogous
notion of simplicity may be relative to common-versus-uncommon
situations and phenomena --- likely a subjective criterion.

Of course, the real desideratum is the design of a time machine,
presumably an amalgam of relativistic, astronomical bodies to produce
large curvature and virtually zero temperature to ensure quantum
coherence and allow regions of negative energy density.

\noindent {\bf Acknowledgement}

The author wishes to thank J. Preskill for his continuing interest,
patient explications, probing questions, and constructive criticism.
This work was supported in part by the U.S. Dept. of Energy under Grant
No. DE-FG03-92-ER40701.

\noindent {\bf Figure Caption}

A class of CTC spacetimes is defined by identifying the like - shaded
time boundaries of two spatial regions at times $t = 0$ and $T$.
Further simplification comes from restricting the space to two points,
$z_1$ and $z_2$, whose respective world histories (broken lines) are
connected and disconnected.

\noindent {\bf References}

\item{1.}  F. Echeverria, G. Klinkhammer, and K.S. Thorne, Phys. Rev.,
{\bf D44}, 1077 (1991).

\item{2.}  D. Deutsch, Phys., Rev., {\bf D44}, 3197 (1991).

\item{3.}  For a recent review with many references, see, e.g., K.S.
Thorne, Ann. N.Y. Ac. Sci., {\bf B631}, 182 (1991).

\item{4.}  P. de Sousa Gerbert and R. Jackiw, Comm. Math. Phys., {\bf
124}, 229 (1989).

\item{5.}  J.L. Friedman, N.J.  Papastamatiou, and J.Z. Simon, Phys.
Rev., {\bf D46}, 4456 (1992).

\item{6.}   D.G. Boulware, Phys. Rev., {\bf D46}, 4421 (1992).

\item{7.}  H.D. Politzer, Phys. Rev., {\bf D46}, 4470 (1992).

\item{8.}  J.B. Hartle, U.C. Santa Barbara Report No. UCSBTH-92-04, 1992
(unpublished).

\item{9.}  R. Geroch and G.T. Horowitz in {\it General Relativity}, S.W.
Hawking and W. Israel, eds., Cambridge University Press, Cambridge
1979.  See also Ref. 7 where non-unitarity of non-relativistic potential
scattering is analyzed in this spacetime.

\item{10.}  F.A. Berezin, The Method of Second Quantization, Academic
Press, New York 1966; pedagogical reviews in L.D. Faddeev and A.A.
Slavnov, {\it Gauge Fields:  Introduction to Quantum Theory},
Benjamin/Cummings Publishing, Reading 1980 and C. Itzykson and J.-B.
Zuber, {\it Quantum Field Theory}, Mc Graw-Hill, New York 1980.

\item{11.}  I.D. Novikov, {\it Evolution of the Universe}, Cambridge
University Press, Cambridge 1983; J. Friedman, M.S. Morris, I.D.
Novikov, F. Echeverria, G. Klinkhammer, K.S. Thorne, and U. Yurtsever,
Phys. Rev., {\bf D42}, 1915 (1990).

\bye